\documentclass[aos,MSNbibl,dvips]{arximspdf}
\usepackage{nicefrac}
\usepackage{mathrsfs,mathbh}
\usepackage{graphicx}

\doi{10.1214/14-AOS1264} 
\volume{42}
\issue{6}
\pubyear{2014}
\firstpage{2469}
\lastpage{2493}
\docsubty{FLA}

\makeatletter
\def\mid{|}
\newcommand{\lleft}{\left}
\newcommand{\rright}{\right}
\def\bfmathh{\bolds}
\def\bfmath{\mathbf}
\newcommand{\eqref}[1]{(\ref{#1})}
\newtheorem{theorem}{Theorem}
\newproclaim{definition}{Definition}
\newtheorem{lemma}[theorem]{Lemma}
\newtheorem{corollary}[theorem]{Corollary}

\def\bfe{\mathbf{e}}
\def\bff{\mathbf{f}}

\def\bfA{\mathbf{A}}
\def\bfB{\mathbf{B}}
\def\bfC{\mathbf{C}}
\def\bfM{\mathbf{M}}
\def\bfW{\mathbf{W}}
\def\bfY{\mathbf{Y}}
\def\bfZ{\mathbf{Z}}

\def\sS{\mathscr{S}}

\def\bbE{\mathbb{E}}
\def\bbR{\mathbb{R}}

\def\cF{{\mathcal F}}
\def\cM{{\mathcal M}}

\def\bbOne{\mathbh{1}}

\def\rank{\operatorname{rank}}

\newcommand{\defeq}{:=}

\newcommand{\fF}{\cF} 
\newcommand{\mF}{\cM} 
\newcommand{\Dom}{\operatorname{Dom}}

\makeatother

\begin{document}
\begin{frontmatter}

\title{Descartes' rule of signs and the identifiability of population
demographic models from genomic variation data\thanksref{T1}}
\runtitle{Identifiability of population demographies}

\begin{aug}
\author[A]{\fnms{Anand}~\snm{Bhaskar}\ead[label=e1]{bhaskar@cs.berkeley.edu}}
\and
\author[B]{\fnms{Yun S.}~\snm{Song}\corref{}\ead[label=e2]{yss@stat.berkeley.edu}}
\runauthor{A. Bhaskar and Y.~S. Song}
\affiliation{University of California, Berkeley}
\address[A]{Computer Science Division\\
University of California, Berkeley\\
Berkeley, California 94720\\
USA\\
\printead{e1}}
\address[B]{Department of Statistics and\\
\quad Computer Science Division\\
University of California, Berkeley\\
Berkeley, California 94720\\
USA\\
\printead{e2}}
\end{aug}
\thankstext{T1}{Supported in part by an NIH Grant R01-GM094402,
and a Packard Fellowship for Science and Engineering.}


\received{\smonth{11} \syear{2013}}
\revised{\smonth{6} \syear{2014}}

%
\begin{abstract}
The sample frequency spectrum (SFS) is a widely-used summary statistic
of genomic variation in a sample of homologous DNA sequences. It
provides a highly efficient dimensional reduction of large-scale
population genomic data and its mathematical dependence on the
underlying population demography is well understood, thus enabling the
development of efficient inference algorithms. However, it has been
recently shown that very different population demographies can actually
generate the same SFS for arbitrarily large sample sizes. Although in
principle this nonidentifiability issue poses a thorny challenge to
statistical inference, the population size functions involved in the
counterexamples are arguably not so biologically realistic. Here, we
revisit this problem and examine the identifiability of demographic
models under the restriction that the population sizes are \emph
{piecewise-defined} where each piece belongs to some family of
biologically-motivated functions. Under this assumption, we prove that
the expected SFS of a sample uniquely determines the underlying
demographic model, provided that the sample is sufficiently large. We
obtain a general bound on the sample size sufficient for
identifiability; the bound depends on the number of pieces in the
demographic model and also on the type of population size function in
each piece. In the cases of piecewise-constant, piecewise-exponential
and piecewise-generalized-exponential models, which are often assumed
in population genomic inferences, we provide explicit formulas for the
bounds as simple functions of the number of pieces. Lastly, we obtain
analogous results for the ``folded'' SFS, which is often used when
there is ambiguity as to which allelic type is ancestral. Our results
are proved using a generalization of Descartes' rule of signs for
polynomials to the Laplace transform of piecewise continuous functions.
\end{abstract}

%

%
\end{frontmatter}
%
\section{Introduction}\label{sec:introduction}

Given a sample of homologous genomic sequences from a large population,
an important inference problem with a wide variety of important
applications is to determine the underlying demography of the
population. The population demography can be used to calibrate null
models of neutral genome evolution in order to find regions under
selection \cite{boyko:2008,lohmueller:2008,williamson:2005}; to
stratify samples in genome-wide association studies \cite
{marchini:2004,campbell:2005,price:2006,pasaniuc:2011}; to date
historical population splits, migrations, admixture and introgression
events \cite{gravel:2011,skoglund:2011,li:2011,lukic:2012,sankararaman:2012,kidd:2012}; and so on. Recently, several
large-sample genome- and exome-sequencing datasets have become
available \cite{coventry:2010,1000G:2010,nelson:2012,tennessen:2012,fu:2012}, shedding new light on patterns of genetic
variation that were not previously observable in smaller datasets. Such
large-sample studies offer an exciting opportunity to infer demography
in unprecedented detail.

One widely-used measure of genetic variation in a set of homologous
genome sequences is the sample frequency spectrum (SFS). For a sample
of size $n$, the SFS counts the proportion of dimorphic (i.e., with
exactly two distinct observed alleles) sites as a function of the
frequency ($\frac{b}{n}$, where $1\leq b\leq n-1$) of the mutant allele
in the sample. The SFS is useful for several reasons.
First, the SFS is a succinct summary of a large sample of genomic
sequences, where the information in $n$ sequences of arbitrary length
can be summarized by just $n - 1$ numbers. This makes the SFS both
mathematically and algorithmically tractable. In particular, since the
SFS ignores linkage information between sites, one can avoid
challenging mathematical and computational issues associated with
rigorously modeling genetic recombination. Furthermore, the statistical
properties of the SFS and their dependence on the population
demographic history are well understood under the coalescent and the
diffusion models of neutral evolution \cite{kimura:1955,fu:1995,griffiths:1998,griffiths:2003,polanski:2003,zivkovic:2011}. This
dependence of the SFS on demography, along with the assumption of free
recombination between sites, has been exploited in several efficient
methods for inferring historical population demography \cite
{marth:2004,gutenkunst:2009,lukic:2011,excoffier:2013}.
Second, the SFS can effectively capture the impact of recent demography
on genetic variation.
Recent large-sample studies \cite{coventry:2010,nelson:2012,tennessen:2012,fu:2012} have consistently shown that there is an
excess of rare polymorphisms compared to the predictions of previously
inferred demographic models, which might be explained by recent rapid
population expansion~\cite{keinan:2012}. Because the leading entries
of the SFS count the rare variants in the sample, one might be able to
use this information to infer demographic events in the recent past at
a much finer resolution than possible using smaller samples.
Third, the SFS also provides a simple way of visualizing the goodness
of fit of a demographic model to data, since one can easily compare the
SFS observed in the data with the SFS predicted by the fitted
demographic model.

While the SFS has algorithmic advantages for demographic inference, it
is believed to suffer from a statistical shortcoming. Specifically,
Myers, Fefferman and
  Patterson \cite{myers:2008} recently showed that even with perfect knowledge of
the \emph{population} frequency spectrum [i.e., the proportion of
polymorphic sites~with population-wide allele frequency in $(x, x+dx)$
for all $x \in(0, 1)$], the historical population size function $\eta
(t)$ as a function of time is not identifiable. Using M\"untz--Sz\'asz
theory, they showed that for any population size function $\eta(t)$,
one can construct arbitrarily many smooth functions $F(t)$ such that
both $\eta(t)$ and $\eta(t) + \alpha F(t)$ generate the same population
frequency spectrum for suitably chosen values of $\alpha$. They also
constructed explicit examples of such functions $\eta(t)$ and $F(t)$.
While this nonidentifiability could pose serious challenges to
demographic inference from frequency spectrum data, the population size
functions involved in their example are arguably unrealistic for
biological populations. In particular, their explicit example involves
a population size function which oscillates at an increasingly higher
frequency as the time parameter approaches the present. Real biological
population sizes can be expected to vary over time in a mathematically
more well-behaved fashion. In particular, populations can be expected
to evolve in discrete units of time, which, when approximated by a
continuous-time model, restricts the frequency of oscillations in the
population size function to be less than the number of generations of
reproduction per unit time. Furthermore, since a population size model
being inferred must have a finite representation for obvious
algorithmic reasons, most previous demographic inference analyses have
focused on inferring population size models that are piecewise-defined
over a restricted class of functions, such as piecewise-constant and
piecewise-exponential models \cite{schaffner:2005,keinan:2007,li:2011,gravel:2011,lukic:2012,nelson:2012,tennessen:2012}.
Motivated by the large number of rare variants observed in several
large-sample sequencing studies, recent works \cite{reppell:2012,reppell:2013} have also focused on more general population growth
models which allow for the population to grow at a faster than
exponential rate. Each piece in such piecewise models has two
parameters that control the rate and acceleration of population growth.
Since these models contain the family of piecewise-constant and
piecewise-exponential population size functions, we refer to them as
piecewise-generalized-exponential models in the remainder of this paper.

In this paper, we revisit the question of demographic model
identifiability under the assumption that the population size is a
\emph
{piecewise-defined} function of time where each piece comes from a
family of biologically-motivated functions, such as the family of
constant or exponential functions.
We also re-examine the assumption that one has access to the
population-wide patterns of polymorphism. In real applications, we do
not expect to know the allele frequency spectrum for an entire
population but rather only the SFS for a randomly drawn finite sample
of individuals. Here, we investigate whether one can learn
piecewise-defined population size functions given perfect knowledge of
the expected SFS for a sufficiently large sample of size $n$. Unlike in
the case of arbitrary continuous population size functions considered
by Myers, Fefferman and
Patterson, the answer to this question is affirmative.
More precisely, we obtain bounds on the sample size $n$ that are
sufficient to distinguish population size functions among piecewise
demographic models with $K$ pieces, where each piece comes from some
family of functions (see Theorems \ref
{thm:unique_eta_general_piecewise_models} and \ref
{thm:equiv_eta_general_models}). Our bound on the sample size can be
expressed as an affine function of the number $K$ of pieces, where the
slope of the function is a measure of the complexity of the family to
which each piece belongs.
In the cases of piecewise-constant, piecewise-exponential and
piecewise-generalized-exponential models, which are often assumed in
population genetic analyses, the slope of this affine function can be
calculated explicitly, as shown in Corollaries \ref
{cor:piecewise_constant}--\ref{cor:piecewise_generalized_exponential}.
We also obtain analogous results for the ``folded'' SFS (see
Theorem~\ref{thm:folded}), a variant of the SFS which circumvents the
ambiguity in
the identity of the ancestral allele type by grouping the polymorphic
sites in a sample according to the sample minor allele frequency.

There are two main technical elements underlying our proofs of the
identifiability results mentioned above. The first step is to show that
the expected SFS of a sample of size $n$ is in bijection with the
Laplace transform of a time-rescaled version of the population size
function evaluated at a particular sequence of $n - 1$ points. This
reduces the problem of identifiability from the SFS to that of
identifiability from the values of the Laplace transform at a fixed set
of points.
The second step relies on a generalization of Descartes' rule of signs
for polynomials to the Laplace transform of general
piecewise-continuous functions. This technique yields an upper bound on
the number of roots of the Laplace transform of a function by the
number of sign changes of the function.
We think that this proof technique based on sign changes might be of
independent interest for proving statistical identifiability results in
other settings.
We also provide an alternate proof of identifiability for
piecewise-constant population models, where the aforementioned second
step is replaced by a linear algebraic argument that has a constructive
flavor. We include this alternate proof in the hope that it could be
used to develop an algebraic inference algorithm for piecewise-constant models.

The remainder of this paper is organized as follows. In Section~\ref
{sec:results}, we introduce the model and notation, and describe our
main results. We also discuss the counterexample of Myers, Fefferman and
  Patterson in light of our findings. The proofs of our results are
provided in Section~\ref{sec:proofs}, and we conclude with a
discussion in
Section~\ref{sec:discussion}.

\section{Main results}\label{sec:results}
Here, we summarize our identifiability results. All proofs are deferred
to Section~\ref{sec:proofs}.

\subsection{Model and notation}

We consider a population evolving according to Kingman's coalescent
\cite{kingman:1982:SPA,kingman:1982:JAP,kingman:1982:EPS} with the
infinite-sites model of mutation \cite{kimura:1969} and selective
neutrality. Under this model, the genome is assumed to be infinite and
every mutation occurs at a different site in the genome that has never
experienced a mutation before. This model is applicable in the regime
where the mutation rate is very low, and hence the probability of
multiple mutations at a given site is vanishingly small. Any
polymorphic site in a sample of sequences is dimorphic under this
model. The population size is assumed to change deterministically with
time and is described by a function $\eta\dvtx\bbR_{\geq0}\to\bbR
_{+}$, such
that the instantaneous coalescence rate between any pair of lineages at
time $t$ is $1/\eta(t)$.

Let $T^{(\eta)}_{n,k}$ denote the time (in coalescent units) while
there are $k$ ancestral lineages for a sample of size $n$ obtained at
time $0$.
Defining $R_{\eta}(t)$ as
\[
R_{\eta}(t):= \int_{0}^{t}
\frac{1}{\eta(x)} \,dx, \label{eq:R}
\]
the expected time $\bbE[T^{(\eta)}_{m,m}]$ to the first coalescence
event for a sample of size $m$ is given by
%
\begin{equation}
\bbE \bigl[T^{(\eta)}_{m,m} \bigr] = \int_0^\infty
t \frac{{m \choose2}}{\eta
(t)} \exp \biggl[ - \pmatrix{m
\cr
2} R_{\eta}(t) \biggr]
\,dt. \label
{eq:c_m_first_exp}
\end{equation}

Following the notation of Myers, Fefferman and
Patterson, define a
time-rescaled version $\widetilde{\eta}$ of the population size
function $\eta$ as
%
\begin{equation}
\widetilde{\eta}(\tau) = \eta \bigl(R_{\eta}^{-1}(\tau) \bigr),
\label
{eq:time_rescaled_N}
\end{equation}
where $\tau\in\bbR_{\geq0}$. The function $\widetilde{\eta}(\tau)$
reparameterizes the population size as a function of the cumulative
rate of coalescence $\tau=R_{\eta}(t)$.
For a given population size function $\widetilde{\eta}$ parameterized
by the total coalescence rate $\tau$, there corresponds a unique
population size function $\eta$ parameterized by time $t$.
Specifically, $\eta(t) = \widetilde{\eta}(S_{\widetilde{\eta
}}^{-1}(t))$, for all $t \in\bbR_{\geq0}$, where $S_{\widetilde
{\eta}}(t)$
is an invertible function given by
\[
S_{\widetilde{\eta}}(t) = \int_{0}^{t} \widetilde{
\eta}(x) \,dx. \label
{eq:inverse_map_time_rescaled_N}
\]
Applying integration by parts to \eqref{eq:c_m_first_exp} and using the
condition that $\bbE[T^{(\eta)}_{m,m}] < \infty$, we have
%
\begin{equation}
\bbE \bigl[T^{(\eta)}_{m,m} \bigr] = \int_0^{\infty}
\exp \biggl[ - \pmatrix{m
\cr
2} R_{\eta}(t) \biggr] \,dt. \label{eq:c_m}
\end{equation}
Furthermore, since $R_{\eta}$ is monotonically increasing and
continuous from $\bbR_{\geq0}$ to $\bbR_{\geq0}$, it is a bijection over
$\bbR_{\geq0}$. For notational convenience, for any interval $I
\subseteq
\bbR_{\geq0}$, we define $R_{\eta}(I)$ to be the interval
%
\[
R_{\eta}(I) = \bigl\{ R_{\eta}(x) \mid x \in I \bigr\}.
\label
{eq:rescale_interval}
\]
By making the substitution $\tau= R_{\eta}(t)$ in \eqref{eq:c_m}
and using \eqref{eq:time_rescaled_N}, we have the following expression
for $\bbE[T^{(\eta)}_{m,m}]$:
%
\begin{equation}
\bbE \bigl[T^{(\eta)}_{m,m} \bigr] = \int_0^{\infty}
\widetilde{\eta}(\tau) \exp \biggl[ - \pmatrix{m
\cr
2} \tau \biggr] \,d\tau\label{eq:c_m_coord_change}.
\end{equation}
Equation \eqref{eq:c_m_coord_change} states that the time to the first
coalescence event for a sample of size $m$ is given by the Laplace
transform of the time-rescaled population size function $\widetilde
{\eta }$ evaluated at the point ${m \choose2}$.
For a sample of size $n$, let $\xi_{n,b}$ denote the probability that a
dimorphic site has $b$ mutant alleles and $n-b$ ancestral alleles. We
refer to $(\xi_{n,1},\ldots,\xi_{n,n-1})$ as the expected sample
frequency spectrum (SFS).

\subsection{Determining the expected times to the first coalescence from the SFS}

The following lemma shows that the expected SFS for a sample of size
$n$ tightly constrains the expected time to the first coalescence event
for all sample sizes $2,\ldots,n$:

\begin{lemma}\label{lem:cm_reconstruct}
Under an arbitrary variable population size model $\{ \eta(t), t\geq
0\}$, suppose $\xi_{n,1},\ldots,\xi_{n,n-1}$ are known and define $c_m
:= \bbE[T^{(\eta)}_{m,m}]$ for $2\leq m \leq n$.
Then, up to a common positive multiplicative constant, the quantities
$c_2,\ldots, c_n$ can be determined uniquely from $\xi_{n,1},\ldots
,\xi
_{n,n-1}$.
\end{lemma}

This implies that the problem of identifying the population size
function $\eta(t)$ from $\xi_{n,1},\ldots,\xi_{n,n-1}$ can be reduced,
up to a multiplicative constant, to the problem of identifying $\eta
(t)$ from $c_2,\ldots, c_n$.

\subsection{Piecewise population size models and sign change complexities}

To state our main result in full generality, we first need a few definitions.

\begin{definition}[($\fF$, family of continuous population size functions)]
A~fa\-mily $\fF$ of continuous population size functions is a set of
positive continuous functions $f\dvtx\bbR_{\geq0}\to\bbR_{+}$ of a
particular
type parameterized by a collection of variables.
\end{definition}

We use $\fF_c$ to denote the family of constant population size
functions; that is, functions of the form $f(t) = \nu$ for all $t$,
where $\nu\in\bbR_{+}$ is the only parameter of the family.
Further, we use $\fF_e$ to denote the family of exponential population
size functions of the form $f(t) = \nu\exp(\beta t)$, where $\nu\in
\bbR_{+}$ and $\beta\in\bbR$ are the parameters of the family.
In human genetics, there has been recent interest \cite
{reppell:2012,reppell:2013} in modeling superexponential growth in the
effective population size via models that generalize exponential growth
by incorporating an additional acceleration parameter $\gamma$. Such
population size functions $f$ satisfy the differential equation $df/dt
= \beta f(t)^{\gamma}$ with initial condition $f(0) = \nu$, where
$\beta\in\bbR, \gamma\in\bbR_{\geq0}$, and $\nu\in\bbR_{+}$.
When $0
\leq\gamma< 1$ (resp., $\gamma> 1$), this represents
superexponential (resp., subexponential) population growth/decline,
while $\gamma= 1$ corresponds to exponential population
growth/decline. We let $\fF_g$ denote the family of such
generalized-exponential population size functions.

\begin{definition}[{[$\mF_K(\fF)$, piecewise models over $\fF$ with at
most $K$ pieces]}]
Given a family $\fF$ of continuous population size functions, a
population size function $\eta(t)$ defined over $\bbR_{\geq0}$ is
said to
be piecewise over $\fF$ with at most $K$ pieces if there exists an
integer $p$, where $1 \leq p \leq K-1$, and a sequence of $p$ time
points $0 < t_1 < \cdots< t_{p} < \infty$ such that for each $1 \leq i
\leq p+1$, there exists a positive continuous function $f_i \in\fF$
such that $\eta(t) = f_i(t - t_{i-1})$ for all $t \in[t_{i-1}, t_i)$.
For convention, we define $t_0 = 0$ and $t_{p+1} = \infty$. Note that
$\eta$ may not be continuous at the change points $t_1,\ldots, t_{p}$.
We use $\mF_K(\fF)$ to denote the space of such piecewise population
size models with at most $K$ pieces, each of which belongs to function
family $\fF$. Illustrated in Figure~\ref{fig:piecewise-eta} is an
example of
piecewise-exponential population size function $\eta\in\mF_K(\fF)$
where $K \geq5$ and~$\fF=\fF_e$.
\end{definition}

\begin{definition}[{[$\sigma(f)$, number of sign changes of a function]}]
For a function $g$ (not necessarily continuous) defined over some
interval $(a, b)$, we say that $t \in(a, b)$ is a sign change point of
$g$ if there exist some $\varepsilon> 0$, $t' \geq t$, and an interval
$(t', t' + \varepsilon) \subseteq(a, b)$ such that:
\begin{longlist}[1.]
\item[1.]$(t - \varepsilon, t) \subseteq(a, b)$,
\item[2.]$g(z) = 0$ for $z \in(t, t')$,

%
\begin{figure}[b]

\includegraphics{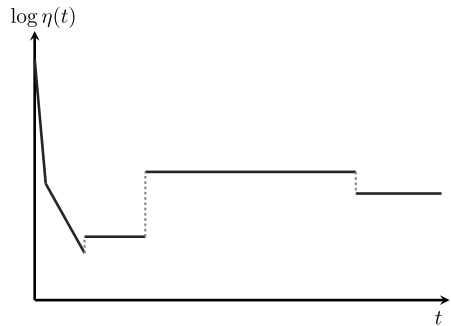}

\caption{A piecewise-exponential population size function $\eta\in
\mF
_K(\fF_e)$, where $K \geq5$. Note that the $y$-axis is in a log scale.
This piecewise-exponential function depicts the historical population
size changes of a European population that was estimated from the SFS
of a sample of $1351$ (diploid) individuals of European ancestry
\protect\cite{tennessen:2012}.}
\label{fig:piecewise-eta}
\end{figure}

%
\begin{figure}

\includegraphics{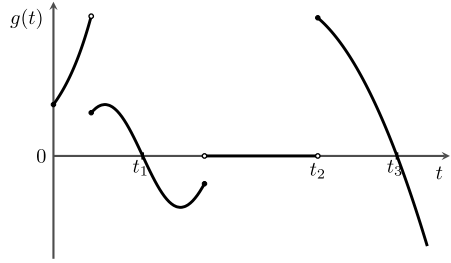}

\caption{Illustration of the sign changes of a function.
For the domain shown, $\sigma(g)=3$ and the sign change points of $g$
are denoted $t_1, t_2$, and $t_3$.}
\label{fig:sign_changes}
\end{figure}

\item[3.]$g(x) g(y) < 0$ for all $x \in(t - \varepsilon, t)$ and $y \in(t',
t' + \varepsilon)$.
\end{longlist}
We define the number $\sigma(g)$ of sign changes of $g$ as the number of
such sign change points in its domain $(a, b)$. See Figure~\ref
{fig:sign_changes} for an illustration.
\end{definition}

Note that the above definition of the number of sign changes counts the
number of times the function $g$ changes value from positive to
negative (and vice versa) while ignoring intervals where it is
identically zero. While the above definition is not restricted to
piecewise continuous functions, we will restrict our attention to such
functions for the remainder of this paper.

%

\begin{definition}[{[$\sS(\fF)$ and $\sS(\mF_K(\fF))$, sign change
complexities]}]
For a family $\fF$ of continuous population size functions, we define
the sign change complexity $\sS(\fF)$ as
%
\begin{eqnarray}\label{eq:sign_change_complexity_fF_exp2}
\sS(\fF) &=& \mathop{\sup_{f_1, f_2 \in\fF, }}_{ a_1, a_2 \in\bbR_{\geq0}} \lleft\{
\sigma(g)\Bigg | %
\begin{array} {l} g(\tau) \defeq \widetilde{f}_1(
\tau- a_1) - \widetilde{f}_2( \tau- a_2)
\mbox{ with domain }
\\
\Dom(g) = \biggl\{ \tau\in\bbR_{\geq0} \bigg| %
\begin{array} {l} \tau- a_1 \in\Dom(\widetilde{f}_1),
\\
\tau- a_2 \in\Dom(\widetilde{f}_2) \end{array}
 \biggr\} \end{array} %
 \rright\}
\nonumber
\\[-8pt]
\\[-8pt]
\nonumber
&=& \mathop{\sup_{f_1, f_2 \in\fF, }}_{ a \in\bbR_{\geq0}} \lleft\{
\sigma(g)\Bigg | %
\begin{array} {l} g(\tau) \defeq \widetilde{f}_1(
\tau) - \widetilde{f}_2(\tau- a) \mbox{ with domain }
\\
\Dom(g) = \biggl\{ \tau\in\bbR_{\geq0} \bigg| %
\begin{array} {l} \tau\in\Dom(\widetilde{f}_1),
\\
\tau- a \in\Dom(\widetilde{f}_2) \end{array} %
 \biggr\}
\end{array} %
 \rright\},
\end{eqnarray}
where $\widetilde{f}_j$ are the time-rescaled versions of $f_j$ as
defined in \eqref{eq:time_rescaled_N}, and $\Dom(\widetilde{f}_j) =
R_{f_j}(\bbR_{\geq0})$ is the domain of $\widetilde{f}_j$.
Similarly, for the space $\mF_K(\fF)$ of piecewise population size
models with at most $K$ pieces over some function family $\fF$, we
define the sign change complexity $\sS(\mF_K(\fF))$ as
%
\[
\sS \bigl(\mF_K(\fF) \bigr) = \sup_{\eta_1, \eta_2 \in\mF_K(\fF)} \bigl\{
\sigma(\widetilde{\eta}_1 - \widetilde{\eta}_2) \bigr\},
\]
where, again, $\widetilde{\eta}_j$ are related to $\eta_j$ as given in
\eqref{eq:time_rescaled_N}.
\end{definition}

The following lemma gives a bound on the sign change complexity of a
model with at most $K$ pieces in terms of the underlying family of
population size functions for each piece.

\begin{lemma}\label{lem:modelsignchanges}
The sign change complexity of the space $\mF_K(\fF)$ of piecewise
models with at most $K$ pieces in a function family $\fF$ is bounded by
the sign change complexity of $\fF$ as
%
\[
\sS \bigl(\mF_K(\fF) \bigr) \leq(2K - 2) + (2K - 1) \sS(\fF).
\]
\end{lemma}

Note that the bound in Lemma~\ref{lem:modelsignchanges} is tight for the
family $\fF_c$ of constant population sizes, for which $\sS(\fF_c) =
0$ and $\sS(\mF_K(\fF_c)) = 2K - 2$.

\subsection{Identifiability results}

Our main results on identifiability will be proved using a
generalization of Descartes' rule of signs for polynomials.

\begin{theorem}[(Descartes' rule of signs for polynomials)]\label
{thm:descartes_rule}
Consider a degree-$n$ polynomial $p(x) = a_0 + a_1 x + \cdots+ a_n
x^n$ with real-valued coefficients $a_i$.
The number of positive real roots (counted with multiplicity) of $p$
is at most the number of sign changes between consecutive nonzero terms
in the sequence $a_0,a_1,\ldots,a_n$.
\end{theorem}


The following theorem generalizes the above classic result to relate
the number of sign changes of a piecewise-continuous function $f$ to
the number of roots of its Laplace transform.

\begin{theorem}[(Generalized Descartes' rule of signs)]\label
{thm:descartes_rule_generalization}
Let $f \dvtx\bbR_{\geq0}\to\bbR$ be a piecewise-continuous
function which
is not identically zero and with a finite number $\sigma(f)$ of sign changes.
Then the function $G(x)$ defined by
%
\begin{equation}
G(x) = \int_0^{\infty} f(t) e^{- t x} \,dt
\label{eq:G_def} 
\end{equation}
has at most $\sigma(f)$ roots in $\bbR$ (counted with multiplicity).
\end{theorem}

The statement of Theorem~\ref{thm:descartes_rule_generalization} and the
proof provided in Section~\ref{sec:proofs} are adapted from
Jameson \cite{jameson:2006}, Lemma~4.5, for our setting. Using Theorem~\ref
{thm:descartes_rule_generalization}, we prove in Section~\ref
{sec:proofs} the
following identifiability theorem for population size function families
with finite sign change complexity.

\begin{theorem}\label{thm:unique_eta_general_models}
For a sample of size $n$, let $\bfmath{c}=(c_2,\ldots,c_n)$, where
$c_m=\bbE
[T^{(\eta)}_{m,m}]$, for $2\leq m \leq n$, defined in \eqref{eq:c_m}.
If $\sS(\fF) < \infty$ and $n \geq\sS(\fF) + 2$, then no two
distinct models $\eta_1,\eta_2\in\fF$ can produce the same
$(c_2,\ldots
,c_n)$. In other words, for $n \geq\sS(\fF) + 2$, the map $\bfmath
{c}\dvtx\fF
\to\bbR_{+}^{n-1}$ is injective.
\end{theorem}

Note that the sample size bound in Theorem~\ref
{thm:unique_eta_general_models} applies to an arbitrary function family
$\fF$ which need not have any special structure. Using Lemma~\ref
{lem:modelsignchanges} for bounding the sign change complexity of
piecewise-defined function families $\mF_K(\fF)$ in terms of the sign
change complexity of the underlying function family $\fF$, we
immediately obtain the following theorem.

\begin{theorem}\label{thm:unique_eta_general_piecewise_models}
For a sample of size $n$, let $\bfmath{c}=(c_2,\ldots,c_n)$, where
$c_m=\bbE
[T^{(\eta)}_{m,m}]$, for $2\leq m \leq n$, defined in \eqref{eq:c_m}.
If $\sS(\fF) < \infty$ and $n \geq2K +(2K-1) \sS(\fF)$,
then the map $\bfmath{c}\dvtx\mF_K(\fF) \to\bbR_{+}^{n-1}$ is injective.
\end{theorem}

Using Theorem~\ref{thm:unique_eta_general_piecewise_models}, it is
simple to
derive identifiability results for piecewise-defined population size
models over several function families $\fF$ that are of biological
interest. In particular, we have the following result for the case of
piecewise-constant models.

\begin{corollary}[{[Identifiability of piecewise-constant population size
models in $\mF_K(\fF_c)$]}]\label{cor:piecewise_constant}
The map $\bfmath{c}\dvtx\mF_K(\fF_c) \to\bbR_{+}^{n-1}$ is
injective if
the sample
size $n \geq2K$.
\end{corollary}

The bound in Corollary~\ref{cor:piecewise_constant} on the sample size
sufficient for identifying piecewise-constant population models is
actually tight, since $\mF_K(\fF_c)$ has $2K - 1$ parameters in
$\bbR_{+}$
and there is no continuous injective function from $\bbR_{+}^{2K-1}$ to
$\bbR^{n-1}$ if $n < 2K$. (This fact can be proved in multiple ways,
such as by the Borsuk--Ulam theorem or the Constant Rank theorem.)
An alternate proof of Corollary~\ref{cor:piecewise_constant} that does not
rely on Theorem~\ref{thm:unique_eta_general_piecewise_models} is also
provided in Section~\ref{sec:proofs}. This alternate proof is based on an
argument from linear algebra, and it might be possible to adapt this
approach to develop an algebraic algorithm for inferring the parameters
of a piecewise-constant population function from the set of expected
first coalescence times $c_m$.

Another class of models often assumed in population genetic analyses
are piecewise-exponential functions, for which we have the following result.

\begin{corollary}[{[Identifiability of piecewise-exponential population
size models in $\mF_K(\fF_e)$]}]\label{cor:piecewise_exponential}
The map $\bfmath{c}\dvtx\mF_K(\fF_e) \to\bbR_{+}^{n-1}$ is
injective if
the sample
size $n \geq4K - 1$.
\end{corollary}

For the generalized-exponential growth models considered by Reppell, Boehnke and
  Z{\"o}llner
  \cite{reppell:2013}, we have the following result.

\begin{corollary}[{[Identifiability of piecewise-generalized-exponential
population size models in $\mF_K(\fF_g)$]}]\label
{cor:piecewise_generalized_exponential}
The map $\bfmath{c}\dvtx\mF_K(\fF_g) \to\bbR_{+}^{n-1}$ is
injective if
the sample
size $n \geq6K - 2$.
\end{corollary}

For the identifiability of piecewise population size models from the
SFS data, we first note the following lemma.

\begin{lemma} \label{lem:rescale}
Consider a piecewise population size function $\eta\in\mF_K(\fF)$.
Consider a sample of size $n \geq2K + (2K-1) \sS(\fF)$ and suppose
the function $\eta$ produces $\bbE[T^{(\eta)}_{m,m}]=c_m$ for $2\leq m
\leq n$.
Then, for every fixed $\kappa\in\bbR_{+}$, there exists a unique piecewise
population size function $\zeta\in\mF_K(\fF)$ with $\bbE[T^{(\zeta
)}_{m,m}]= \kappa c_m$ for $2\leq m \leq n$. Furthermore, this
population size function $\zeta$ is given by $\zeta(t) = \kappa\eta
(t/\kappa)$.
\end{lemma}

Given two models $\eta,\zeta\in\mF_K$, we say that $\eta$ and
$\zeta$
are equivalent, and write $\eta\sim\zeta$, if they are related by a
rescaling of change points and population sizes as described in
Lemma~\ref{lem:rescale}.
Let $[\eta]$ denote the equivalence class of population size functions
that contain $\eta$, and let $\mF_K(\fF)/ {\sim}{}= \{ [\eta] \mid
\eta\in\mF_K(\fF)\}$ be the set of equivalence classes for the
equivalence relation $\sim$.
Then, combining Lemma~\ref{lem:cm_reconstruct}, Theorem~\ref
{thm:unique_eta_general_piecewise_models} and Lemma~\ref{lem:rescale},
we obtain the following theorem.

\begin{theorem} \label{thm:equiv_eta_general_models}
If $\sS(\fF) < \infty$ and $n\geq2K + (2K - 1) \sS(\fF)$, then,
for each expected SFS $(\xi_{n,1},\ldots,\xi_{n,n-1})$, there exists a
unique equivalence class $[\eta]$ of models in $\mF_K(\fF)/ {\sim}$
consistent with $(\xi_{n,1},\ldots,\xi_{n,n-1})$.
\end{theorem}

\subsection{Extension to the folded frequency spectrum}

To generate the SFS from genomic sequence data, one needs to know the
identities of the ancestral and mutant alleles at each site.
To avoid this problem, a commonly employed strategy in population
genetic inference involves ``folding'' the SFS.
More precisely, for a sample of size $n$, the $i$th entry of the folded
SFS $\bfmathh{\chi} = (\chi_{n,1}, \ldots, \chi_{\lfloor{n}/{2}
\rfloor})$ is defined by
%
\[
\chi_{n,i} = \frac{\xi_{n,i} + \xi_{n,n-i}}{1+\delta_{i,n-i}},
\]
where $1 \leq i \leq\lfloor{n}/{2} \rfloor$. In particular,
$\chi_{n,i}$ is the proportion of polymorphic sites that have $i$
copies of the minor allele. For any sample size $n$, since $\bfmath
{\chi
}$ is a vector of approximately half the dimension as $\bfmathh{\xi}$,
we might expect to require roughly twice as many samples to recover the
demographic model from $\bfmathh{\chi}$ compared to $\bfmathh{\xi}$. This
is indeed the case. Given the folded SFS $\bfmathh{\chi}$, the following
theorem establishes a sufficiency condition on the sample size for
identifying demographic models in $\mF_K(\fF)$.

\begin{theorem}\label{thm:folded}
If $\sS(\fF) < \infty$ and $n\geq2 (2K - 1) (1 + \sS(\fF))$,
then, for each expected folded SFS $\bfmathh{\chi} = (\chi
_{n,1},\ldots
,\chi_{n,{\lfloor{n}/{2} \rfloor}})$, there exists a unique
equivalence class $[\eta]$ of models in $\mF_K(\fF)/ {\sim}$
consistent with $\bfmathh{\chi}$.
\end{theorem}

\subsection{The counterexample of Myers, Fefferman and Patterson}\label{sec:myers_discussion}

Myers, Fefferman and
  Patterson \cite{myers:2008} provided an explicit counterexample to the
identifiability of population size models from the allelic frequency
spectrum. In our notation, they provided two time-rescaled population
size functions $\widetilde{\eta}_1$ and $\widetilde{\eta}_2$ given by
%
\begin{eqnarray*}
\widetilde{\eta}_1(\tau) &=& N,
\\
\widetilde{\eta}_2(\tau) &=& N \bigl(1 - 9 F(\tau) \bigr),
\end{eqnarray*}
where $N$ is an arbitrary positive constant, and the function $F$ is
given by the convolution
%
\[
F(\tau) = \int_0^{\tau} f_0(\tau- u)
f_1(u) \,du,
\]
where $f_0$ and $f_1$ are given by
%
\begin{eqnarray*}
f_0(\tau) &=& \exp \bigl(-1 / \tau^2 \bigr),
\\
f_1(\tau) &= &\frac{\cos(\pi^2/\tau) \exp(-\tau/8)}{\sqrt{\tau}}.
\end{eqnarray*}
Both functions $f_1$ and $F$ have increasingly frequent oscillations as
$\tau\downarrow0$ so that $\sigma(\widetilde{\eta}_1 - \widetilde
{\eta }_2) = \sigma(F) = \infty$. This is why Theorem~\ref
{thm:unique_eta_general_models} does not apply to this example.
Indeed, by an argument using the Laplace transforms of $f_1$ and $F$,
Myers, Fefferman and
  Patterson showed that the function $G(x)$ defined in
\eqref{eq:G_def} in terms of $F$ has roots at $-{m \choose2}$ for each
$m \geq2$.

\section{Proofs}\label{sec:proofs}

We now provide proofs of the results presented earlier.

\begin{pf*}{Proof of Lemma~\ref{lem:cm_reconstruct}}
In the coalescent for a sample of size $n$, let $\gamma_{n,b}$ denote
the total expected branch length subtending $b$ leaves, for $1 \leq b
\leq n - 1$. Then $\xi_{n,b} = \gamma_{n,b}/\sum_{k=1}^{n-1} \gamma
_{n,k}$, which implies that there exists a positive constant $\kappa$
such that $\gamma_{n,b} = \kappa\xi_{n,b}$ for all $1 \leq b \leq n -
1$. We now prove that $c_2,\ldots,c_n$ can be determined uniquely from
$\gamma_{n,1},\ldots,\gamma_{n,n-1}$.

Let $\phi_{n,k} = \bbE[T^{(\eta)}_{n,k}]$. Then, by a result of
Griffiths and Tavar\'e \cite{griffiths:1998},
%
\begin{equation}
\gamma_{n,b} = \sum_{k = 2}^{n - b + 1} k
\frac{{n - b - 1 \choose k -
2}}{{n - 1 \choose k - 1}} \phi_{n,k}, \label{eq:tau_T}
\end{equation}
for $1 \leq b \leq n - 1$.
The system of equations \eqref{eq:tau_T} can be rewritten succinctly
as a linear system
%
\[
\bfmathh{\gamma} = \bfM\bfmathh{\phi},
\]
where $\bfmathh{\gamma}=(\gamma_{n,1},\ldots,\gamma_{n,n-1})$,
$\bfmath
{\phi}=(\phi_{n,2},\ldots,\phi_{n,n})$, and $\bfM= (m_{bk})$ with
$m_{bk} = k {{n - b - 1 \choose k - 2}} / {{n - 1 \choose k - 1}}$, for
$1\leq b \leq n-1$ and $2\leq k \leq n$.
The matrix $\bfM$ is upper-left triangular since ${n - b - 1 \choose k
- 2} = 0$ if $k > n - b + 1$, and the anti-diagonal entries are $\frac
{k}{{n - 1 \choose k - 1}} > 0$. Hence, $\det(\bfM) \neq0$ and $\bfM$
is therefore invertible. Thus, given $\bfmathh{\gamma}$, we can
determine $\bfmathh{\phi}$ uniquely as $\bfM^{-1}\bfmathh{\gamma}$.

Let $\psi_{n,k}= \sum_{j=k}^n {\bbE[T^{(\eta)}_{n,j}]}$. Then,
defining $\psi_{n,n+1} \defeq0$, observe that $\psi_{n,k} = \phi_{n,k}
+ \psi_{n,k + 1}$ for $2 \leq k \leq n$. This implies that $\psi
_{n,2},\ldots,\psi_{n,n}$ can be determined uniquely from $\phi
_{n,2},\ldots,\phi_{n,n}$.
Polanski, Bobrowski and
  Kimmel \cite{polanski:2003} showed that $\psi_{n,k}$ can be written as
%
\begin{equation}
\psi_{n,k} = \sum_{m=k}^{n}
a_{km} c_m, \label{eq:psi_cm}
\end{equation}
where $a_{km}$, for $k \leq m \leq n$, are given by
%
\[
a_{km} = \frac{\prod_{l=k, l \neq m}^n{l \choose2}}{\prod_{l=k, l
\neq m}^n  [{l \choose 2} -{m \choose 2} ]},
\]
and $c_m=\bbE[T^{(\eta)}_{m,m}]$, shown in \eqref{eq:c_m}. Again, the
system of equations \eqref{eq:psi_cm} can be written as a triangular
linear system
%
\[
\bfmathh{\psi} = \bfA\bfmath{c},
\]
where $\bfmathh{\psi}=(\psi_{n,2},\ldots,\psi_{n,n})$, $\bfmath
{c}=(c_2,\ldots
,c_n)$, and $\bfA= (a_{km})$, for $2\leq k,m\leq n$.
Note that $\bfA$ is an upper triangular matrix since $a_{km} \defeq0$
if $m < k$. Since $\bfA$ has nonzero entries on its diagonal, $\bfA
^{-1}$ exists, and $\bfmath{c}$ can be determined uniquely as $\bfA
^{-1}\bfmathh{\psi}$.
\end{pf*}

\begin{pf*}{Proof of Lemma~\ref{lem:modelsignchanges}}
Given a pair of piecewise population size functions $\eta_1, \eta_2
\in\mF_K(\fF)$, let $\widetilde{\eta}_1$ and $\widetilde{\eta
}_2$ be
their respective time-rescaled versions, defined by \eqref
{eq:time_rescaled_N}. Let $0 < t^{(1)}_1 < \cdots< t^{(1)}_{p_1} <
\infty$, where $0 \leq p_1 \leq K - 1$ (resp., $0 < t^{(2)}_1 <
\cdots
< t^{(2)}_{p_2} < \infty$, where $0 \leq p_2 \leq K - 1$) be the change
points of the pieces of $\eta_1$ (resp., $\eta_2$).
We define $t^{(1)}_{0}=t^{(2)}_{0}=0$ and
$t^{(1)}_{p_1+1}=t^{(2)}_{p_2+1}=\infty$.
The change points of $\widetilde{\eta}_1$ are given by $R_{\eta
_1}(t^{(1)}_i)$, where $1 \leq i \leq p_1$, while the change points of
$\widetilde{\eta}_2$ are given by $R_{\eta_2}(t^{(2)}_i)$, where
$1 \leq i \leq p_2$.
Let $0 < \tau_1 < \cdots< \tau_p < \infty$ be the union of the change
points of $\widetilde{\eta}_1$ and $\widetilde{\eta}_2$, where $0
\leq
p \leq p_1 + p_2$. For convention, let $\tau_0 = 0$ and $\tau_{p + 1} =
\infty$.

Consider the piece $(\tau_i, \tau_{i+1})$ for $0 \leq i \leq p$. Let
$I_1=(t^{(1)}_{k}, t^{(1)}_{k+1})$, where $0 \leq k \leq p_1$, and $I_2
= (t^{(2)}_{l}, t^{(2)}_{l+1})$, where $0\leq l \leq p_2$, be the
pieces of the original population size functions $\eta_1$ and $\eta_2$,
respectively, such that $(\tau_i, \tau_{i+1}) \subseteq R_{\eta
_1}(I_1)$ and $(\tau_i, \tau_{i+1}) \subseteq R_{\eta_2}(I_2)$.
Since $\eta_1 \in\mF_K(\fF)$, there exists a function $f_1 \in\fF$
such that $\eta_1(t) = f_1(t - t^{(1)}_{k})$ for all $t \in I_1$. Then,
for all $\tau\in R_{\eta_1}(I_1)$,
%
\begin{eqnarray}\label{eq:eta1_f1_rel}
\widetilde{\eta}_1(\tau) &=& \eta_1
\bigl(R_{\eta _1}^{-1}(\tau) \bigr) 
= f_1
\bigl(R_{\eta_1}^{-1}(\tau) - t^{(1)}_{k}
\bigr)\nonumber
\\
&=& \widetilde{f}_1 \bigl(R_{f_1} \bigl(R_{\eta _1}^{-1}(
\tau) - t^{(1)}_{k} \bigr) \bigr)
\\
&=& \widetilde{f}_1 \bigl(\tau- R_{\eta_1}
\bigl(t^{(1)}_{k} \bigr) \bigr). \nonumber
\end{eqnarray}
Similarly, there exists some function $f_2 \in\fF$ such that, for all
$\tau\in R_{\eta_2}(I_2)$,
%
\begin{equation}
\widetilde{\eta}_2(\tau) = \widetilde{f}_2 \bigl(\tau-
R_{\eta
_2} \bigl(t^{(2)}_{l} \bigr) \bigr).
\label{eq:eta2_f2_rel}
\end{equation}
Using \eqref{eq:eta1_f1_rel} and \eqref{eq:eta2_f2_rel}, we see that
the number of sign change points of $\widetilde{\eta}_1 - \widetilde
{\eta}_2$ in the piece $(\tau_i, \tau_{i+1})$ is at most the number of
sign change points of $\widetilde{f}_1(\tau- R_{\eta
_1}(t^{(1)}_{k})) - \widetilde{f}_2(\tau- R_{\eta _2}(t^{(2)}_{l}))$
for $\tau\in(\tau_i, \tau_{i+1})$. Hence, by
\eqref
{eq:sign_change_complexity_fF_exp2}, it follows that within each piece
$(\tau_i, \tau_{i+1})$ for $0 \leq i \leq p$, $\widetilde{\eta}_1 -
\widetilde{\eta}_2$ has at most $\sS(\fF)$ sign change points.
Also, the point $\tau_{i+1}$ itself could be a sign change point in the
interval between the last sign change point in piece $(\tau_i, \tau
_{i+1})$ and the first sign change point in piece $(\tau_{i+1}, \tau
_{i+2})$ where $0 \leq i \leq p - 1$. These are all the possible sign
change points of $\widetilde{\eta}_1 - \widetilde{\eta}_2$.
Hence,
%
\begin{eqnarray}\label{eq:ineq_sch_etas}
\sigma(\widetilde{\eta}_1 - \widetilde{\eta}_2) &\leq& p
+ (p + 1) \sS(\fF)
\nonumber
\\
&\leq&(p_1 + p_2) + (p_1 + p_2
+ 1) \sS(\fF)
\\
&\leq&(2K - 2) + (2K - 1) \sS(\fF). \nonumber
\end{eqnarray}
Since \eqref{eq:ineq_sch_etas} holds for all $\eta_1, \eta_2 \in\mF
_K(\fF)$, the lemma follows.
\end{pf*}

\begin{pf*}{Proof of Theorem~\ref{thm:descartes_rule_generalization}}
The proof is by induction on the number of sign changes of $f$. If $f$
has zero sign changes, then without loss of generality, $f(t) \geq0$
for $t \in(0, \infty)$ and $f(t) > 0$ for some interval $(a, b)
\subseteq(0, \infty)$. Hence, $G(x) > 0$ for all $x$, and the base
case holds. Suppose $f$ has $m + 1$ sign change points $t_0, \ldots,
t_m$, where $m \geq0$. Note that $G(x)$ and $F(x) = e^{t_0 x} G(x)$
have the same real-valued roots (with multiplicity) since $e^{t_0 x} >
0$ for all $x \in\bbR$. $F'(x)$ is given by
%
\[
F'(x) = \frac{d}{dx} \biggl( \int_0^{\infty}
f(t) e^{- (t - t_0) x} \,dt \biggr) = \int_0^{\infty}
(t_0 - t) f(t) e^{- (t - t_0) x} \,dt,
\]
where the interchange of the differential and integral operators in the
second equality is justified by the Leibniz integral rule because $f$
is piecewise continuous over $\bbR_{\geq0}$, and both $f(t) e^{- (t -
t_0) x}$ and $\frac{d}{dx} (f(t) e^{- (t - t_0) x})$ are jointly
continuous over $(p_i, p_{i+1}) \times(-\infty, \infty)$ for each
piece $(p_i, p_{i+1})$ over which $f$ is continuous.
Note that the set of sign change points of $(t_0 - t) f(t)$ is $\{t_1,
\ldots, t_m \}$. Hence, $(t_0 - t) f(t)$ has only $m$ sign changes. By
the induction hypothesis, $F'$ has at most $m$ real-valued roots. By
Rolle's theorem, the number of real-valued roots of $F$ is at most one
more than the number of real-valued roots of $F'$. Hence, $F$ has at
most $m + 1$ real-valued roots, implying that $G$ has at most $m + 1$
real-valued roots.
\end{pf*}

\begin{pf*}{Proof of Theorem~\ref{thm:unique_eta_general_models}}
Suppose there exist two distinct population size functions $\eta_1,
\eta_2 \in\fF$ that produce exactly the same $c_m$ for all $2 \leq m
\leq n$.
From~\eqref{eq:c_m_coord_change}, we have that
%
\begin{equation}
\int_0^{\infty} \bigl(\widetilde{
\eta}_1(\tau) - \widetilde{\eta }_2(\tau) \bigr)
e^{- {m \choose2} \tau}\,d\tau= 0 \label{eq:roots_G}
\end{equation}
for $2 \leq m \leq n$.
If we define the function $G(x)$ as
%
\[
G(x) = \int_0^{\infty} \bigl(\widetilde{
\eta}_1(\tau) - \widetilde {\eta }_2(\tau) \bigr)
e^{- x \tau} \,d\tau,
\]
then from \eqref{eq:roots_G}, we see that ${m \choose2}$ is a root of
$G(x)$ for $2 \leq m \leq n$, and hence, $G$~has at least $n - 1$ roots.
Applying Theorem~\ref{thm:descartes_rule_generalization} to the piecewise
continuous function $\widetilde{\eta}_1 - \widetilde{\eta}_2$, we see
that $G$ can have at most $\sigma(\widetilde{\eta}_1 - \widetilde
{\eta }_2)$ roots. Taking the supremum over all population size functions
$\eta_1$ and $\eta_2$ in $\fF$, we see that $G$ can have at most
$\sS(\fF)$ roots. Hence, if $n - 1 > \sS(\fF)$, we get a contradiction.
This implies that if $n \geq\sS(\fF) + 2$, no two distinct
population size functions in $\fF$ can produce the same $(c_2,\ldots,c_n)$.
\end{pf*}

\begin{pf*}{Proof of Corollary~\ref{cor:piecewise_constant}}
As remarked after Lemma~\ref{lem:modelsignchanges}, for the constant
population size function family $\fF_c$, $\sS(\fF_c) = 0$. Hence, by
Theorem~\ref{thm:unique_eta_general_piecewise_models}, if $n \geq2K$, the
map $\bfmath{c}\dvtx\mF_K(\fF_c) \to\bbR_{+}^{n-1}$ is injective.
\end{pf*}

%
\begin{pf*}{An alternate proof of Corollary~\ref{cor:piecewise_constant}
based on linear algebra}
Let $n \geq2K$, and suppose there exist two distinct models $\eta
^{(1)}, \eta^{(2)} \in\mF_K(\fF_c)$ that produce exactly the same
$c_m$ for all $2 \leq m \leq n$. Let $\widetilde{\eta}^{(1)}$ and
$\widetilde{\eta}^{(2)}$ denote the time-rescaled versions of $\eta
^{(1)}$ and $\eta^{(2)}$, respectively, as in \eqref
{eq:time_rescaled_N}. Since $\eta^{(j)}$ is piecewise constant with at
most $K$ pieces, $\widetilde{\eta}^{(j)}$ is also piecewise constant
with the same number of pieces as $\eta^{(j)}$, and $\eta^{(1)} \neq
\eta^{(2)}$ implies $\widetilde{\eta}^{(1)} \neq\widetilde{\eta}^{(2)}$.
Therefore, $\widetilde{\Delta}:= \widetilde{\eta}^{(1)} -
\widetilde{\eta}^{(2)}$ is a piecewise-constant function over $[0,
\infty)$ with
$p$ pieces, where $1 \leq p \leq2K - 1$, and $\widetilde{\Delta}$ is
not identically zero. Let $\tau_1 < \cdots< \tau_{p-1}$ denote the
change points of $\widetilde{\Delta}$, and define $\tau_0=0$ and
$\tau
_p = \infty$. Suppose $\widetilde{\Delta}(\tau) = \delta_i \in
\bbR$
for all $\tau\in[\tau_{i-1}, \tau_i)$, where $1 \leq i \leq p$.
Since $\widetilde{\eta}^{(1)}$ and $\widetilde{\eta}^{(2)}$ produce
the same $c_m$ for all $2 \leq m \leq n$, we know that $\widetilde
{\Delta}$ satisfies
%
\begin{equation}
\int_{0}^{\infty} \widetilde{\Delta}(\tau)
e^{- {m \choose2} \tau} \,d\tau= 0, \label{eq:integral_system}
\end{equation}
for all $2 \leq m \leq n$.
Substituting the definition of $\widetilde{\Delta}$ into \eqref
{eq:integral_system} and multiplying by ${m \choose2}$, we obtain
%
\begin{equation}
\sum_{i=1}^{p} \delta_i
\bigl[ e^{- {m \choose2} \tau_{i-1}} - e^{- {m
\choose2} \tau_i} \bigr] = 0, \label{eq:linear_system}
\end{equation}
for $2 \leq m \leq n$. This defines a linear system $\bfA\bfmath
{\delta
} = \bfmath{0}$, where $\bfmathh{\delta}=(\delta_1,\ldots,\delta
_p)$ and
$\bfA= (a_{mi})$ is an $(n-1) \times p$ matrix with $a_{mi}:= e^{-{m
\choose2} \tau_{i-1}} - e^{-{m \choose2} \tau_i}$ for $2 \leq m
\leq
n$ and $1 \leq i \leq p$.

Let $\bfB= (b_{mi})$ be the $(n-1) \times p$ matrix formed from $\bfA
$ such that the $i$th column of $\bfB$ is the sum of columns $i,
i+1, \ldots, p$ of $\bfA$. Defining $\alpha_i = e^{- \tau_{i-1}}$, note
that $b_{mi} = \alpha_i^{{m \choose2}}$ for $2 \leq m \leq n$ and $1
\leq i \leq p$.
Now, consider the $p\times p$ submatrix $\bfC$ of $\bfB$ consisting of
the first $p$ rows of $\bfB$.
Since $\alpha_1 > \alpha_2 > \cdots> \alpha_p > 0$, note that $\bfC$
is a generalized Vandermonde matrix, which implies $\det(\bfC) \neq0$
\cite{gantmacher:2000}, Chapter~XIII, Section~8.
Hence, $\rank(\bfB) = p$. The rank of $\bfA$ is invariant under
elementary column operations and, therefore, $\rank(\bfA)=\rank(\bfB
) =p$.
Therefore, the kernel of $\bfA$ is trivial, and the only solution to
\eqref{eq:linear_system} is $\delta_1 = \delta_2 = \cdots= \delta
_p =
0$, which contradicts our assumption that $\widetilde{\Delta} =
\widetilde{\eta}^{(1)} - \widetilde{\eta}^{(2)} \not\equiv0$.
\end{pf*}

\begin{pf*}{Proof of Corollary~\ref{cor:piecewise_exponential}}
Let $f_1, f_2 \in\fF_e$ be given by
%
\begin{eqnarray*}
f_1(t) &=& \nu_1 \exp(\beta_1 t),
\\
f_2(t) &= &\nu_2 \exp(\beta_2 t),
\end{eqnarray*}
where $t \in\bbR_{\geq0}$, $\nu_1, \nu_2 \in\bbR_{+}$ and $\beta_1,
\beta_2 \in\bbR$.
Then, for $i=1,2$, the time-rescaled function $\widetilde{f}_i$ is
given by
%
\begin{equation}
\widetilde{f}_i(\tau) = \frac{\nu_i}{1 - \nu_i \beta_i \tau}, \label
{eq:eta1_rescaled_exp_pop}
\end{equation}
for $\tau\in\Dom(\widetilde{f_i}) = R_{f_i}(\bbR_{\geq0}) =
[0, \frac{1}{\nu_i \beta_i})$.
From \eqref{eq:eta1_rescaled_exp_pop}, it can be seen that $\widetilde
{f}_1$ and $\widetilde{f}_2$ are continuous in their domains.
Furthermore, for any given $a\in\bbR_{\geq0}$, there is at most one
$\tau$, where $\tau\in\Dom(\widetilde{f_1})$ and $\tau-a\in\Dom
(\widetilde{f_2})$, such that $g(\tau)\defeq\widetilde{f}_1(\tau) -
\widetilde{f}_2(\tau-a) = 0$, implying $\sigma(g) \leq1$.
By the definition of sign change complexity in \eqref
{eq:sign_change_complexity_fF_exp2}, it then follows that $\sS(\fF_e)
\leq1$ for the exponential population family $\fF_e$.
Hence, applying Theorem~\ref{thm:unique_eta_general_piecewise_models}, we
conclude that $n \geq4K - 1$ suffices for the map $\bfmath{c}\dvtx
\mF
_K(\fF_e)
\to\bbR_{+}^{n-1}$ to be injective.
\end{pf*}

\begin{pf*}{Proof of Corollary~\ref{cor:piecewise_generalized_exponential}}
Let $f_1, f_2 \in\fF_g$ be generalized-exponential functions which
satisfy the following differential equations and initial conditions:
%
\begin{eqnarray*}
\frac{df_i}{dt} &=& \beta_i f_i(t)^{\gamma_i},
\\
f_i(0) &=& \nu_i, \qquad i \in\{1, 2\},
\end{eqnarray*}
where $\nu_i \in\bbR_{+}$, $\beta_i \in\bbR$ and $\gamma_i \in
\bbR_{\geq0}$.
The solutions for $f_i$ are given by
%
\[
f_i(t) = %
\cases{ \nu_i \exp(
\beta_i t), &\quad $\gamma_i = 1$, \vspace*{2pt}
\cr
 \bigl[ \nu_{i}^{1-\gamma_i} +
\beta_i t (1 - \gamma_i) \bigr]^{{1}/{(1-\gamma_i)}}, &\quad  $
\gamma_i \neq1$.} %
\]
It can be shown that the time-rescaled population size functions
$\widetilde{f_i}$ are given by
%
\begin{equation}
\widetilde{f_i}(\tau) = %
\cases{ \nu_i
\exp(\beta_i \tau), &\quad $\beta_i = 0 \mbox{ or }
\gamma_i = 0$,\vspace*{2pt}
\cr
\bigl(\nu_i^{-\gamma_i}
- \beta_i \gamma_i \tau \bigr)^{-{1}/{\gamma
_i}}, &\quad $
\beta_i \neq0 \mbox{ and } \gamma_i > 0$.}
\label{eq:rescalef_forms}
\end{equation}
In order to obtain an upper bound on $\sS(\fF_g)$, we consider the
following three cases depending on the functional form of $\widetilde
{f_1}$ and $\widetilde{f_2}$ in \eqref{eq:rescalef_forms}:
\begin{itemize}
\item\emph{Case} 1: $\widetilde{f_1}(\tau) = \nu_1 \exp(\beta_1
\tau)$
and $\widetilde{f_2}(\tau) = \nu_2 \exp(\beta_2 \tau)$. Since
$\widetilde{f_1}$ and $\widetilde{f_2}$ are continuous functions of
$\tau$, the number of sign changes of $g(\tau)\defeq\widetilde
{f_1}(\tau
) - \widetilde{f_2}(\tau-a)$ is at most the number of roots of
$g(\tau
)$. Taking the logarithm of $\widetilde{f_1}(\tau)$ and $\widetilde
{f_2}(\tau- a)$, it is easy to see that $g(\tau)$ has at most one root
for any $a\in\bbR_{\geq0}$. Hence, $\sigma(g) \leq1$.

\item\emph{Case} 2: $\widetilde{f_1}$ and $\widetilde{f_2}$ have
different functional forms. Suppose
$\widetilde{f_1}(\tau) =\break \nu_1 \exp(\beta_1 \tau)$ and
$\widetilde{f_2}(\tau) = (\nu_2^{-\gamma_2} - \beta_2 \gamma_2
\tau)^{-
{1}/{\gamma_2}}$. For any\vspace*{1pt} $a_1, a_2 \in\bbR_{\geq0}$ such that $\tau
- a_i
\in\Dom(\widetilde{f_i})$, the number of sign changes of $g(\tau
)\defeq
\widetilde{f_1}(\tau- a_1) - \widetilde{f_2}(\tau- a_2)$ is at most
the number of roots of $g(\tau)$.
By raising $\widetilde{f_1}(\tau- a_1)$ and $\widetilde{f_2}(\tau-
a_2)$ to the power of $- \gamma_2$, we see that the number of roots of
$g(\tau)$ is the number of solutions to
%
\begin{equation}
\mu_1 \exp(-\gamma_2 \beta_1 \tau) =
\mu_2^{-\gamma_2} - \beta_2 \gamma_2 \tau,
\label{eq:rescalef_case2}
\end{equation}
where $\mu_1 = \nu_1 \exp(\gamma_2 \beta_1 a_1)$ and $\mu_2 =
(\nu
_2^{-\gamma_2} + \beta_2 \gamma_2 a_2)^{-{1}/{\gamma_2}}$.
Equation \eqref
{eq:rescalef_case2} represents the intersection of an exponential
function with a line and has at most 2 solutions for $\tau$. Hence,
$\sigma(g) \leq2$.

\item\emph{Case} 3: $\widetilde{f_i}(\tau) = (\nu_i^{-\gamma_i} -
\beta
_i \gamma_i \tau)^{-{1}/{\gamma_i}}$ for $i = 1, 2$.
Let $g(\tau)\defeq\widetilde{f_1}(\tau) - \widetilde{f_2}(\tau-a)$
where $a \in\bbR_{\geq0}$ such that $\tau- a \in\Dom(\widetilde{f_2})$.
Since $g$ is a continuous function, the number of sign changes of
$g(\tau)$ in $\bbR_{\geq0}$ is bounded by the number of distinct positive
roots of $g(\tau)$. The number of distinct positive roots of $g$ is the
number of distinct positive solutions $\tau$ to
%
\[
\bigl(\nu_1^{-\gamma_1} - \beta_1
\gamma_1 \tau \bigr)^{-{1}/{\gamma
_1}} = \bigl(\nu_2^{-\gamma_2}
- \beta_2 \gamma_2 (\tau- a) \bigr)^{-{1}/{\gamma_2}},
\]
which is also the number of distinct positive solutions to
%
\begin{equation}
\bigl(\nu_1^{-\gamma_1} - \beta_1
\gamma_1 \tau \bigr)^{{\gamma_2}/{\gamma
_1}} = \nu_2^{-\gamma_2}
- \beta_2 \gamma_2 (\tau- a). \label
{eq:rescalef_case3}
\end{equation}
Let $x \defeq\widetilde{f_1}(\tau)^{-\gamma_1} = \nu_1^{-\gamma
_1} -
\beta_1 \gamma_1 \tau$. Since $\widetilde{f_1}$ is a time-rescaled
population size function, $x > 0$ when $\tau\in\bbR_{\geq0}$. Since
$\beta
_i \neq0$ and $\gamma_i > 0$, \eqref{eq:rescalef_case3} can be
rewritten as
\[
x^{{\gamma_2}/{\gamma_1}} + A x + B = 0, \label{eq:rescalef_case3_xeqn}
\]
where $A = - \frac{\beta_2 \gamma_2}{\beta_1 \gamma_1}$ and $B =
\frac
{\beta_2 \gamma_2}{\beta_1 \gamma_1} \nu_1^{-\gamma_1} - \nu
_2^{-\gamma
_2} - \beta_2 \gamma_2 a$. Letting $h(x) \defeq x^{{\gamma
_2}/{\gamma_1}} + A x + B$, the number of distinct positive solutions
for $\tau$ in \eqref{eq:rescalef_case3} is at most the number of
distinct positive roots for the generalized polynomial $h$.
For any real-valued function $g(x)$ possessing infinitely many
derivatives and any interval $I \subseteq\bbR$, let $Z(g, I)$ the
number of zeroes of $g$ contained in $I$, counted with multiplicity. By
a consequence of Rolle's theorem \cite{jameson:2006}, Proposition~2.1,
$Z(g, I) \leq Z(g', I) + 1$. Observing that $h'(x) = \frac{\gamma
_2}{\gamma_1} x^{{\gamma_2}/{\gamma_1} - 1} + A$ has at most one
root in $\bbR_{+}$, $Z(h, \bbR_{+}) \leq Z(h', \bbR_{+}) + 1 \leq
2$. Hence, the
number of distinct positive solutions $\tau$ to \eqref
{eq:rescalef_case3} is at most 2, and $\sigma(g) \leq2$.
\end{itemize}

From the definition of sign change complexity in \eqref
{eq:sign_change_complexity_fF_exp2} and the bound on $\sigma(g)$ in the
three cases above, it follows that $\sS(\fF_g) \leq2$ for the
generalized-exponential population family $\fF_g$.
Hence, applying Theorem~\ref{thm:unique_eta_general_piecewise_models}, we
conclude that $n \geq6K - 2$ suffices for the map $\bfmath{c}\dvtx
\mF
_K(\fF_g)
\to\bbR_{+}^{n-1}$ to be injective.
\end{pf*}

\begin{pf*}{Proof of Lemma~\ref{lem:rescale}}
For the population size function $\zeta(t)$ defined by $\zeta(t) =
\kappa\eta(t/\kappa)$, note that $R_{\zeta}(t)$ is given by
\[
R_{\zeta}(t) = \int_0^t
\frac{1}{\zeta(x)} \,dx = \int_0^t
\frac
{1}{\kappa\eta(x/\kappa)} \,dx = \int_0^{t/\kappa}
\frac{1}{\eta
(x)} \,dx = R_{\eta}(t/\kappa).
\]
$\bbE[T^{(\zeta)}_{m,m}]$ is then given by
%
\begin{eqnarray*}
\bbE \bigl[T^{(\zeta)}_{m,m} \bigr] &=& \int_0^{\infty}
\exp \biggl[ - \pmatrix{m
\cr
2} R_{\eta} \biggl(\frac{t}{\kappa}
\biggr) \biggr] \,dt
\\
&=& \kappa\int_0^{\infty} \exp \biggl[ - \pmatrix{m
\cr
2} R_{\eta }(t) \biggr] \,dt
\\
&=& \kappa\bbE \bigl[T^{(\eta)}_{m,m} \bigr].
\end{eqnarray*}
Since $n \geq2K + (2K-1) \sS(\fF)$, by Theorem~\ref
{thm:unique_eta_general_piecewise_models}, $\zeta$ is the unique
population size function in $\mF_K(\fF)$ with $\bbE[T^{(\zeta)}_{m,m}]=
\kappa c_m$ for $2\leq m \leq n$.
\end{pf*}

To prove Theorem~\ref{thm:folded}, we first need a lemma that characterizes
a certain symmetry property of the invertible matrix that relates the
genealogical quantities $\bfmathh{\gamma}$ and $\bfmath{c}$
introduced in the
proof of Lemma~\ref{lem:cm_reconstruct}.

%
\begin{lemma}\label{lem:W_odd_cols}
For a sample of size $n$, let $W$ be the $(n-1) \times(n-1)$
invertible matrix such that $\gamma_{n,b} = \sum_{m=2}^n W_{b,m} c_m$,
where $\gamma_{n,b}$ is the total expected branch length subtending $b$
leaves and $c_m = \bbE[T^{(\eta)}_{m,m}]$. Then, for every $b$ and $m$,
where $1 \leq b \leq n - 1$ and $2 \leq m \leq n$, we have the
following identities:
%
\begin{eqnarray*}
W_{b,m} + W_{n-b,m} &=& 0\qquad \mbox{if } m \mbox{ is odd},
\\
W_{b,m} - W_{n-b,m} &=& 0\qquad \mbox{if } m \mbox{ is even}.
\end{eqnarray*}
\end{lemma}
%

\begin{pf}
From the proof of Lemma~\ref{lem:cm_reconstruct}, it can be seen that
the matrix $\bfW$ is the product of 3 matrices whose entries are
explicitly given combinatorial expressions. However, using Zeilberger's
algorithm \cite{petkovsek:1996}, Polanski and Kimmel \cite{polanski:2003-1}, equations
(13)--(15), also derived the following recurrence relation
for the entries of $\bfW$:
%
\begin{eqnarray}\label{eq:W_recur}
W_{b,2} &=& \frac{6}{(n+1)},
\nonumber
\\
W_{b,3} &= &30 \frac{(n-2b)}{(n+1)(n+2)},
\\
W_{b,m+2} &=& f(n,m) W_{b,m} + g(n,m) (n-2b) W_{b,m+1},
\nonumber
\end{eqnarray}
where $f(n,m)$ and $g(n,m)$ are rational functions of $n$ and $m$ given by
\begin{eqnarray*}
f(n,m) &=& -\frac{(1+m)(3+2m)(n-m)}{m(2m-1)(n+m+1)},
\\
g(n,m) &=& \frac{(3+2m)}{m(n+m+1)}.
\end{eqnarray*}
It will be easy to prove our lemma by induction on $m$ using \eqref
{eq:W_recur}. The base cases are easy to check:
\begin{eqnarray*}
W_{b,2} - W_{n-b,2} &=& 0,
\\
W_{b,3} + W_{n-b,3} &=& 30 \frac{(n - 2b) + (n - 2(n-b))}{(n+1)(n+2)} = 0.
\end{eqnarray*}
Using \eqref{eq:W_recur}, we see that if $m$ is odd,
\begin{eqnarray*}
&&W_{b,m+2} + W_{n-b,m+2}
\\
&&\qquad= f(n,m) (W_{b,m} + W_{n-b,m})
\\
&&\qquad\quad{} + g(n,m) \bigl\{(n-2b) W_{b,m+1} + \bigl[n-2(n-b) \bigr]
W_{n-b,m+1} \bigr\}
\\
&&\qquad= f(n,m) (W_{b,m} + W_{n-b,m}) + g(n,m) (n-2b)
(W_{b,m+1} - W_{n-b,m+1})
\\
&&\qquad= 0,
\end{eqnarray*}
where the last equality follows from the induction hypothesis which
implies $W_{b,m} + W_{n-b,m}=0$ and $W_{b,m+1} - W_{n-b,m+1}=0$.
Similarly, if $m$ is even,
\begin{eqnarray*}
&&W_{b,m+2} - W_{n-b,m+2}
\\
&&\qquad= f(n,m) (W_{b,m} - W_{n-b,m})
\\
&&\qquad\quad{} + g(n,m) \bigl\{ (n-2b) W_{b,m+1} - \bigl[n-2(n-b) \bigr]
W_{n-b,m+1} \bigr\}
\\
&&\qquad= f(n,m) (W_{b,m} - W_{n-b,m}) + g(n,m) (n-2b)
(W_{b,m+1} + W_{n-b,m+1})
\\
&&\qquad= 0,
\end{eqnarray*}
where again the last equality follows from the induction hypothesis.\vadjust{\goodbreak}
\end{pf}

\begin{pf*}{Proof of Theorem~\ref{thm:folded}}
For a sample of size $n$ in the coalescent, let $\gamma_{n,b}$ be the
total expected branch length subtending $b$ leaves, for $1 \leq b \leq
n - 1$. Then there exists a positive constant $\kappa$ such that
%
\begin{equation}
\frac{\gamma_{n,d} + \gamma_{n,n-d}}{1 + \delta_{d,n-d}} = \kappa \chi _{n,d}, \label{eq:constant_folded_spectrum}
\end{equation}
for all $1 \leq d \leq{\lfloor{n}/{2} \rfloor}$. Let $f_{n,d}
= \frac{\gamma_{n,d} + \gamma_{n,n-d}}{1 + \delta_{d,n-d}}$. The
relationship between $\bff= (f_{n,1}, \ldots, f_{n,{\lfloor{n}/{2} \rfloor}})$ and $\bfmathh{\gamma} = (\gamma_{n,1}, \ldots,
\gamma
_{n,n-1})$ can be described by the linear equation
%
\[
\bff= \bfZ\bfmathh{\gamma},
\]
where $\bfZ$ is an $\lfloor{n}/{2} \rfloor\times(n - 1)$
matrix with entries given by
%
\[
Z_{dj} = %
\cases{ 1, &\quad $\mbox{if } j = d \mbox{ or } j = n
- d$, \vspace*{2pt}
\cr
0, & \quad $\mbox{otherwise}$, } %
\]
for $1 \leq d \leq\lfloor{n}/{2} \rfloor$ and $1 \leq j
\leq
n - 1$. Hence, $\dim(\ker(\bfZ)) = \lfloor{(n-1)}/{2}
\rfloor$.

From Lemma~\ref{lem:cm_reconstruct}, we know that $\bfmathh{\gamma}$ and
$\bfmath{c}=(c_2,\ldots,c_n)$ are related as $\bfmathh{\gamma} =
\bfW\bfmath
{c}$, where $\bfW= (W_{b,m})$ is an $(n-1) \times(n-1)$ invertible
matrix, where $1\leq b \leq n-1$ and $2\leq m \leq n$. Hence,
%
\begin{equation}
\bff= \bfY\bfmath{c}, \label{eq:folded_spectrum_linear_system}
\end{equation}
where $\bfY\defeq\bfZ\bfW$. Since $Y_{b,m} = W_{b,m} + W_{n-b,m}$,
we know from Lemma~\ref{lem:W_odd_cols} that $Y_{b,m} = 0$ for all odd
values of $m$. Therefore, every other column of the matrix $\bfY$ is
zero. This implies that $\operatorname{span}(\{ \bfe_3, \bfe_5,
\ldots, \bfe
_{n-\bbOne\{n \mbox{ even}\}} \}) \subseteq\ker(\bfY)$, where
$\bfe
_{i}$ is an $(n-1)$-dimensional unit vector defined as $\bfe_{i}=
(e_{i,2},\ldots,e_{i,n})$, with $e_{i,i}=1$ and $e_{i,j}=0$ for $i\neq
j$. Note that $n-\bbOne\{n \mbox{ even}\}=2\lfloor{(n-1)}/{2}\rfloor+ 1$ and $\dim(\operatorname{span} (\{ \bfe_3,
\bfe_5, \ldots
, \bfe_{2\lfloor{(n-1)}/{2}\rfloor+ 1}\})) = \lfloor
{(n-1)}/{2}\rfloor$.
Now, since $\bfW$ is invertible, $\dim(\ker(\bfY)) = \dim(\ker
(\bfZ
\bfW)) = \dim(\ker(\bfZ)) = \lfloor{(n-1)}/{2}\rfloor$.\break
Therefore,
%
\begin{equation}
\ker(\bfY) = \operatorname{span} \bigl(\{ \bfe_3,
\bfe_5, \ldots , \bfe _{2\lfloor{(n-1)}/{2}\rfloor+ 1} \} \bigr). \label{eq:ker_Y}
\end{equation}

Suppose there exist two distinct models $\eta_1, \eta_2 \in\mF
_K(\fF
)$ that produce the same folded SFS $\bff$. Let $\bfmath{c}^{(1)}$
and $\bfmath{c}
^{(2)}$ be the vector of genealogical quantities for models $\eta_1$
and $\eta_2$, respectively, where $c^{(1)}_m=\bbE[T^{(\eta_1)}_{m,m}]$
and $c^{(2)}_m=\bbE[T^{(\eta_2)}_{m,m}]$, $2 \leq m \leq n$.
From \eqref{eq:folded_spectrum_linear_system}, we know that $\bfmath{c}
^{(1)} - \bfmath{c}^{(2)} \in\ker(\bfY)$. Using \eqref{eq:ker_Y},
$c^{(1)}_m
- c^{(2)}_m$ can be written as
%
\begin{equation}
c^{(1)}_m - c^{(2)}_m = \sum
_{l=1}^{\lfloor
{(n-1)}/{2}\rfloor
} \alpha_l
e_{2l+1,m}, \label{eq:c_m_ker_Y}
\end{equation}
for some $\alpha_l \in\bbR$.
Since $e_{ij}=0$ for $i\neq j$, \eqref{eq:c_m_ker_Y} implies that
$c^{(1)}_m - c^{(2)}_m = 0$ for all even values of $m$, where $2 \leq m
\leq n$.
Now applying a similar argument as in the proof of Theorem~\ref
{thm:unique_eta_general_piecewise_models} to $c^{(1)}_m - c^{(2)}_m$
for even values of $m$, we conclude that if $ \lceil{(n -
1)}/{2}  \rceil> (2K - 2) + (2K - 1) \sS(\fF)$, then no two
distinct models $\eta_1, \eta_2 \in\mF_K(\fF)$ can produce the same
$\bff$. This implies that a sample size $n \geq2 (2K - 1) (1 + \sS
(\fF))$ suffices for identifying the population size function in $\mF
_K(\fF)$ from the folded SFS $\bff$, and the conclusion of the theorem
follows from \eqref{eq:constant_folded_spectrum} and Lemma~\ref{lem:rescale}.
\end{pf*}

\section{Discussion}\label{sec:discussion}

In human genetics, several large-sample datasets have recently become
available, with sample sizes on the order of several thousands to tens
of thousands of individuals \cite{1000G:2010,coventry:2010,fu:2012,nelson:2012,tennessen:2012}.
The patterns of polymorphism observed in these datasets deviate
significantly from that expected under a constant population size, and
there has been much interest in inferring recent and ancient human
demographic changes that might explain these deviations \cite
{gravel:2011,li:2011,lukic:2012}.
Clearly, model identifiability is an important prerequisite for such
statistical inference problems.
In this paper, we have obtained mathematically rigorous identifiability
results for demographic inference by showing that piecewise-defined
population size functions over a wide class of function families are
completely determined by the SFS, provided that the sample is
sufficiently large.
Furthermore, we have provided explicit bounds on the sample sizes that
are sufficient for identifying such piecewise population size
functions. These bounds depend on the number of pieces and the
functional type of each piece. For piecewise-constant population size
models, which have been extensively applied in demographic inference
studies, our bounds are tight. We have also given analogous results for
identifiability from the folded SFS, a variant of the SFS that is
oblivious to the identities of the ancestral and mutant alleles.

\begin{figure}

\includegraphics{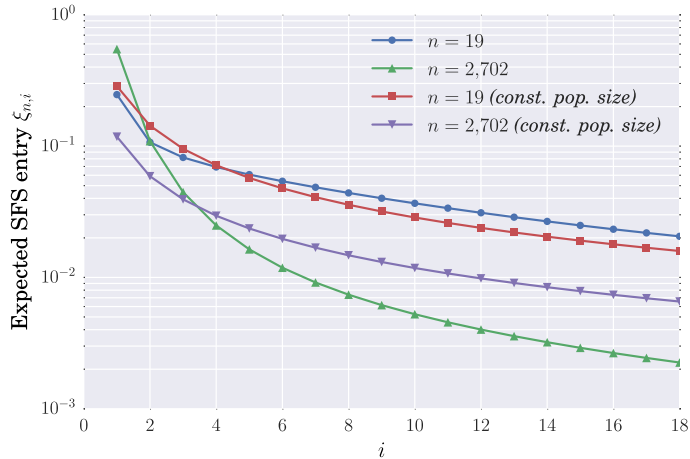}

\caption{The leading entries of the expected SFS $\bfmathh{\xi}_n$ for
a piecewise-exponential population size model inferred b
Tennessen et~al. \protect\cite{tennessen:2012}. This demographic model, shown (up to scaling) in
Figure~\protect\ref{fig:piecewise-eta}, was fitted using the observed
SFS from a sample of
$1351$ (diploid) individuals of European ancestry~\protect\cite
{tennessen:2012}. The blue plot is the expected SFS for $n = 19$, which
matches the sample size bound in Corollary~\protect\ref
{cor:piecewise_exponential}
for identifying piecewise-exponential models with up to 5 pieces, while
the green plot is the first 18 entries of the expected SFS for $n =
2702$ ($1351$ diploids). The red and purple plots are the
expected SFS for $n = 19$ and $n=2702$, respectively, for a constant
population size function.}
\label{fig:expectedSFSTennessenModel}
\end{figure}

Recent large-sample sequencing studies have consistently found a
substantially higher fraction of rare variants compared to the
predictions of the coalescent with a constant population size, even in
regions of the genome that are believed to have evolved neutrally
\cite
{gazave:2013}. Keinan and Clark \cite{keinan:2012} suggested that recent rapid
expansion of the population has given rise to variants which are
private to single individuals in the population, and that this
signature of population expansion is particularly apparent now due to
the larger sample sizes involved in sequencing studies.
We illustrate this point with a specific example. The blue plot in
Figure~\ref{fig:expectedSFSTennessenModel} shows the expected SFS for
a sample of
size $n=19$ under the piecewise-exponential population size history
with 5 epochs recently inferred by Tennessen et~al. \cite{tennessen:2012} and
illustrated in Figure~\ref{fig:piecewise-eta}. (Note that $n=19$ is the
sample size bound given by Corollary~\ref{cor:piecewise_exponential} for
identifying piecewise-exponential models with up to 5 pieces.) The red
plot in Figure~\ref{fig:expectedSFSTennessenModel} shows the expected
SFS for
the same sample size under a constant population size model.
For this small sample size, the two expected frequency spectra are very
similar despite the large difference in demographic models, indicating
the difficulty of accurately recovering the details of recent
exponential population growth using small-sample data.
In contrast, for a much larger sample of size $n=2702$, which
corresponds to the actual sample size for Tennessen et~al.'s
data, the expected frequency spectra under the two demographic models
mentioned above are considerably more different; see the green and
purple plots in Figure~\ref{fig:expectedSFSTennessenModel}.

On the other hand, our identifiability results show that \emph{perfect}
data (i.e., the \emph{exact} expected SFS) from even a small sample
size of $n = 4K - 1$ are sufficient to uniquely identify a
piecewise-exponential model with $K$ pieces. This gap between
theoretical identifiability and practical inference needs to be better
addressed through robustness results that can account for the finite
genome length, which limits the resolution to which the expected SFS of
a random sample can be estimated. Our identifiability results apply in
the limit that the genome length is infinite, which allows one to
estimate the entries of the expected SFS exactly. On the other hand, a
finite length genome does not permit exact estimation of the expected
SFS, which can make it difficult in practice to resolve the details of
ancient demographic events even if the sample size is large. This is
because population size changes sufficiently far back in the past are
likely to have only a marginal effect on the SFS since the individuals
in the sample are highly likely to have found a common ancestor by such
ancient times.


Our work suggests several interesting avenues for future research. An
important problem is to understand the sensitivity of the SFS to
perturbations in the demographic parameters. A related problem is
quantifying the extent to which errors in estimating the expected SFS
from a finite amount of data affect the parameter estimates in inferred
demographic models.

It would also be interesting to consider the possibility of developing
an algebraic algorithm for demographic inference that closely mimics
the linear algebraic proof of Corollary~\ref{cor:piecewise_constant} provided
in Section~\ref{sec:proofs}. For example, using a sample of size $K +
1$, one
could consider inferring a piecewise-constant model with $K$ pieces,
with one piece for each of the most recent $K - 1$ generations and
another piece for the population size further back in time. (Here, we
are considering a restricted class of piecewise-constant population
size functions with fixed change points, so the minimum sample size
needed for distinguishing such models using the SFS is $K+1$ rather
than $2K$.) Such an algebraic algorithm could provide a more principled
way of inferring demographic parameters, compared to existing inference
methods that rely on optimization procedures which lack theoretical
guarantees for functions with multiple local optima.

In our work, we focused on the identifiability of demography from the
expected SFS data.
However, if one were to use the complete sequence data or other summary
statistics such as the length distribution of shared haplotype tracts,
it might be possible to uniquely identify the demography using even
smaller sample sizes than that needed when using only the SFS. Indeed,
several demographic inference methods have been developed to infer
historical population size changes from such data using anywhere from a
pair of genomic sequences \cite{li:2011,harris:2013,palamara:2012}
to tens of such sequences \cite{sheehan:2013}, and it is important to
theoretically characterize the power and limitations of both the data
and the inference methods.

\section*{Acknowledgments}
We thank Graham Coop, Noah Mattoon, Aylwyn Scally and anonymous
reviewers for their comments on our work.
We also thank the generous support of the Simons Institute for the
Theory of Computing. The final version of this work was completed while
the authors were participating in the 2014 program on ``Evolutionary
Biology and the Theory of Computing.''


%

\printaddresses

\end{document}